\documentclass[onecolumn,tighten,times,fleqn]{aastex631}
\usepackage{color}
\usepackage{multirow}
\usepackage{chngcntr}
\usepackage{mathtools}
\usepackage{amsmath}
\usepackage{amsfonts}
\usepackage{amssymb}
\usepackage[utf8]{inputenc}  
\usepackage{graphicx}        
\usepackage{perpage}         
\usepackage{bm}
\usepackage{hyperref}
\usepackage{url}
\MakePerPage{footnote}

\graphicspath{{./}{figures/}}

\makeatletter

\newcommand{\Rmnum}[1]{\expandafter\@slowromancap\romannumeral #1@}
\makeatother

\newcommand{\ha}{H$\alpha$}
\newcommand{\hb}{H$\beta$}
\newcommand{\oiii}{[O\,{\sc iii}]}
\newcommand{\nii}{[N\,{\sc ii}]}
\newcommand{\sii}{[S\,{\sc ii}]}
\newcommand{\oii}{[O\,{\sc ii}]}
\newcommand{\oi}{[O\,{\sc i}]}

\newcommand{\kms}{\mathrm{km~s^{-1}}}

\shorttitle{DESI--SDSS Aperture Bias}
\shortauthors{Liu \& Rong}

\begin{document}

\title{A Direct DESI--SDSS Two-Fibre Test of Aperture Bias in Optical Emission-Line Diagnostics}

\author{Shihong Liu}
\affiliation{Department of Astronomy, University of Science and Technology of China, Hefei, Anhui 230026, China}
\affiliation{School of Astronomy and Space Sciences, University of Science and Technology of China, Hefei 230026, Anhui, China}

\correspondingauthor{Yu Rong}
\email{rongyua@ustc.edu.cn}

\author{Yu Rong$^*$}
\affiliation{Department of Astronomy, University of Science and Technology of China, Hefei, Anhui 230026, China}
\affiliation{School of Astronomy and Space Sciences, University of Science and Technology of China, Hefei 230026, Anhui, China}




\begin{abstract}
Fixed angular apertures sample different physical regions of nearby galaxies and can therefore bias optical emission-line diagnostics. We use 20,545 high-confidence emission-line galaxies observed by both the Dark Energy Spectroscopic Instrument (DESI) and the Sloan Digital Sky Survey (SDSS) as a two-fibre experiment, in which the DESI 1.5 arcsec fibre is nested within the SDSS 3 arcsec fibre. To isolate aperture effects from line-measurement systematics, we refit every spectrum using the same eMILES stellar-continuum and emission-line model while preserving the wavelength-dependent line-spread function of each spectrum. This common refit yields 20,200 quality-controlled matched galaxies. Relative to SDSS, the smaller DESI aperture produces small but highly significant offsets: $\Delta\log({\rm N2})=+0.0067\pm0.0002$, $\Delta\log({\rm S2})=-0.0148\pm0.0003$, $\Delta\log({\rm O3})=-0.0246\pm0.0006$, $\Delta{\rm O3N2}=-0.0327\pm0.0007$, and $\Delta\log({\rm H}\alpha/{\rm H}\beta)=-0.0195\pm0.0003$, where $\Delta$ denotes DESI minus SDSS. These offsets persist in Baldwin--Phillips--Terlevich-selected star-forming galaxies and in a near-concentric subset. Their amplitudes vary across the redshift distribution and retain a measurable dependence on apparent fibre coverage after redshift is controlled, showing that redshift and relative aperture coverage jointly modulate the aperture terms. We conclude that, for high-S/N emission-line galaxies, DESI--SDSS comparisons carry measurable, diagnostic-dependent fibre-aperture terms even when continuum modelling, emission-line fitting, and spectral-resolution treatment are held fixed.
\end{abstract}

\keywords{methods: statistics \-- galaxies: ISM \-- galaxies: abundances \-- techniques: spectroscopic \-- surveys} 


\section{Introduction}
Optical nebular emission lines are among the most widely used probes of galaxy evolution. Ratios involving \oii, \hb, \oiii, \ha, \nii, \sii, and \oi\ are used to infer gas-phase metallicity, dust attenuation, ionization parameter, electron density, star-formation activity, and excitation by active galactic nuclei (AGN) or shocks \citep{Baldwin1981,Kewley2001,Kauffmann2003,Brinchmann2004,Tremonti2004,Kewley2019}. Empirical and theoretical strong-line calibrations, including N2, O3N2, R23, O32, and related indices, underpin measurements of the mass--metallicity relation, abundance gradients, and environmental trends \citep{Pettini2004,Kewley2008,Marino2013,Curti2017,Maiolino2019}. Large spectroscopic samples are now routinely used to test differential metallicities and line-ratio residuals associated with star-formation rate, local environment, and position within galaxies, often at the $\lesssim0.05$ dex level \citep{Mannucci2010,Ellison2009,Curti2020,Belfiore2017b}. Survey-dependent systematics of only a few hundredths of a dex can therefore become astrophysically important.

A long-recognized source of such systematics is aperture bias. A fixed angular fibre does not measure a galaxy-integrated spectrum; rather, it samples only the region enclosed by the fibre. This matters because galaxies exhibit radial gradients in stellar age, dust attenuation, star-formation surface density, metallicity, ionization parameter, diffuse ionized gas fraction, and nuclear excitation. The issue was already evident in the Sloan Digital Sky Survey (SDSS) era, when the 3 arcsec fibre enabled enormous statistical samples but sampled different physical fractions of galaxies as a function of redshift and size \citep{York2000,Hopkins2003,Brinchmann2004,Kewley2005}. Dedicated comparisons between nuclear and integrated spectroscopy showed that aperture coverage affects star-formation rates, reddening, and metallicity estimates \citep{Moustakas2006,Kewley2005}. Calar Alto Legacy Integral Field Area (CALIFA) aperture simulations further demonstrated systematic growth curves with aperture radius for the Balmer decrement and standard strong-line diagnostics \citep{IglesiasParamo2016}. Integral-field surveys, including CALIFA \citep{Sanchez2012}, the Sydney--AAO Multi-object Integral-field spectrograph (SAMI) \citep{Croom2012}, and Mapping Nearby Galaxies at Apache Point Observatory (MaNGA) \citep{Bundy2015}, have since shown directly that nebular line ratios and stellar-population properties vary substantially within galaxies \citep{Sanchez2014,Belfiore2017}.

The Dark Energy Spectroscopic Instrument (DESI) brings this issue into a new regime. DESI is delivering millions of spectra over a large sky area, with a 1.5 arcsec-diameter fibre and broad wavelength coverage suitable for nebular diagnostics in low-redshift galaxies \citep{Aghamousa2016,DESI2022,Silber2023,DESI2026}. DESI emission-line products are already enabling large-sample measurements of galaxy properties across successive data releases: early DESI data have been used to identify extremely metal-poor galaxies and to characterize their abundance patterns \citep{Zou2024,Zinchenko2024}, while the DR1 products have enabled a larger census of extremely metal-poor systems \citep{Sui2026}. Other applications include studies of DESI [O\,II] emission-line profiles and their links to star formation and morphology \citep{Lan2024}, dwarf active galactic nucleus candidates \citep{Pucha2025}, the [O\,II]--[S\,II] electron-density offset \citep{Rong26b}, and the structural dependence of cool-gas outflows \citep{Rong26a}. However, much of the empirical calibration framework and many comparison samples are inherited from SDSS, whose fibres are 3 arcsec in diameter. A direct DESI--SDSS comparison is therefore a prerequisite for placing DESI line-ratio measurements on the same footing as SDSS-based calibrations and for assessing whether mixed-survey samples carry subtle aperture-dependent offsets. The DESI and SDSS spectra also have different, wavelength-dependent spectral-resolution functions, so a meaningful comparison must distinguish aperture effects from line-spread-function effects.

The overlap between DESI and SDSS offers a clean two-aperture experiment. For nearly concentric observations, DESI samples the central 1.5 arcsec-diameter region, while SDSS samples a larger 3 arcsec-diameter aperture around the same galaxy. The same galaxy can therefore serve as its own control. This experiment complements integral-field-unit aperture studies: IFU surveys reveal the spatial origin of aperture effects in relatively small samples \citep{Sanchez2014,Belfiore2017}, whereas the DESI--SDSS overlap directly tests the net bias present in the catalogues and apertures used by large spectroscopic surveys.

The primary objective of this work is to measure, with one spectral-fitting pipeline, the diagnostic-dependent aperture terms that arise when the same low-redshift galaxy is observed through the nested DESI and SDSS fibres. Its novelty is the direct large-sample DESI DR1--SDSS comparison: it isolates the aperture term from differences in the public survey pipelines and tests explicitly how that term changes with redshift and apparent galaxy size. This is needed both for cross-survey measurements and for redshift-dependent aperture effects within an individual fibre survey.

In this paper, we measure empirical aperture terms in standard emission-line diagnostics. The central analysis is a same-pipeline spectral-refitting experiment applied to all matched spectra, so that DESI and SDSS are compared using the same continuum model, emission-line model, and line-spread-function (LSF) treatment. Public catalogue fluxes are used only for the initial high-confidence line selection and for context. We define \citep{Pettini2004,Marino2013}
\begin{align}
{\rm N2} &\equiv [{\rm N\,II}]\lambda6583/{\rm H}\alpha, \\
{\rm S2} &\equiv ([{\rm S\,II}]\lambda6716+[{\rm S\,II}]\lambda6731)/{\rm H}\alpha, \\
{\rm O3} &\equiv [{\rm O\,III}]\lambda5007/{\rm H}\beta, \\
{\rm O3N2} &\equiv \log{\rm O3}-\log{\rm N2}.
\end{align}
The Balmer decrement is ${\rm H}\alpha/{\rm H}\beta$. The \sii\ density-sensitive ratio is $[{\rm S\,II}]\lambda6716/[{\rm S\,II}]\lambda6731$. The Baldwin--Phillips--Terlevich (BPT) diagram is the plane of $\log{\rm O3}$ versus $\log{\rm N2}$ used to distinguish star formation from harder ionizing sources \citep{Baldwin1981,Kauffmann2003}. Unless otherwise stated, all offsets are defined as $\Delta X=X_{\rm DESI}-X_{\rm SDSS}$.

\section{Materials and Methods}
\label{sec:data}

\subsection{Matched DESI--SDSS spectra}
We match galaxies with both DESI Data Release 1 (DR1) and SDSS Data Release 7 (DR7) optical spectra \citep{York2000,Abazajian2009,DESI2026}. We require the DESI and SDSS fibre centres to be separated by less than 0.75 arcsec. Because the DESI fibre radius is 0.75 arcsec and the SDSS fibre radius is 1.5 arcsec, this criterion ensures that the DESI fibre footprint lies within the SDSS fibre footprint. We also require the redshifts to agree within a line-of-sight velocity difference of $|\Delta v|<100~\kms$, where $\Delta v\equiv c(z_{\rm DESI}-z_{\rm SDSS})/(1+z_{\rm SDSS})$ and $c$ is the speed of light.

For the main analysis, we use the DESI emission-line catalogue of \citet{Zou2024} and the Max Planck Institute for Astrophysics--Johns Hopkins University (MPA--JHU) SDSS DR7 emission-line catalogue \citep{Brinchmann2004,Tremonti2004} only to pre-select galaxies with robust measurements of the diagnostics studied here. We require the six individual diagnostic lines \hb, \oiii$\lambda5007$, \ha, \nii$\lambda6583$, and \sii$\lambda\lambda6716,6731$ to have signal-to-noise ratio (S/N) $>5$ in both catalogues. This defines a deliberately high-S/N emission-line sample rather than a complete census of the DESI--SDSS overlap. The selection yields 20,545 galaxies for which both the SDSS spectrum and the local DESI coadded spectrum are available. Catalogue measurements are not used as the final line fluxes: all ratios reported below are measured from our common spectral refit. The sample is designed to test aperture terms in the BPT, Balmer-decrement, \sii-density, N2, S2, O3, and O3N2 diagnostics.

\begin{figure*}
\centering
\includegraphics[width=0.80\textwidth]{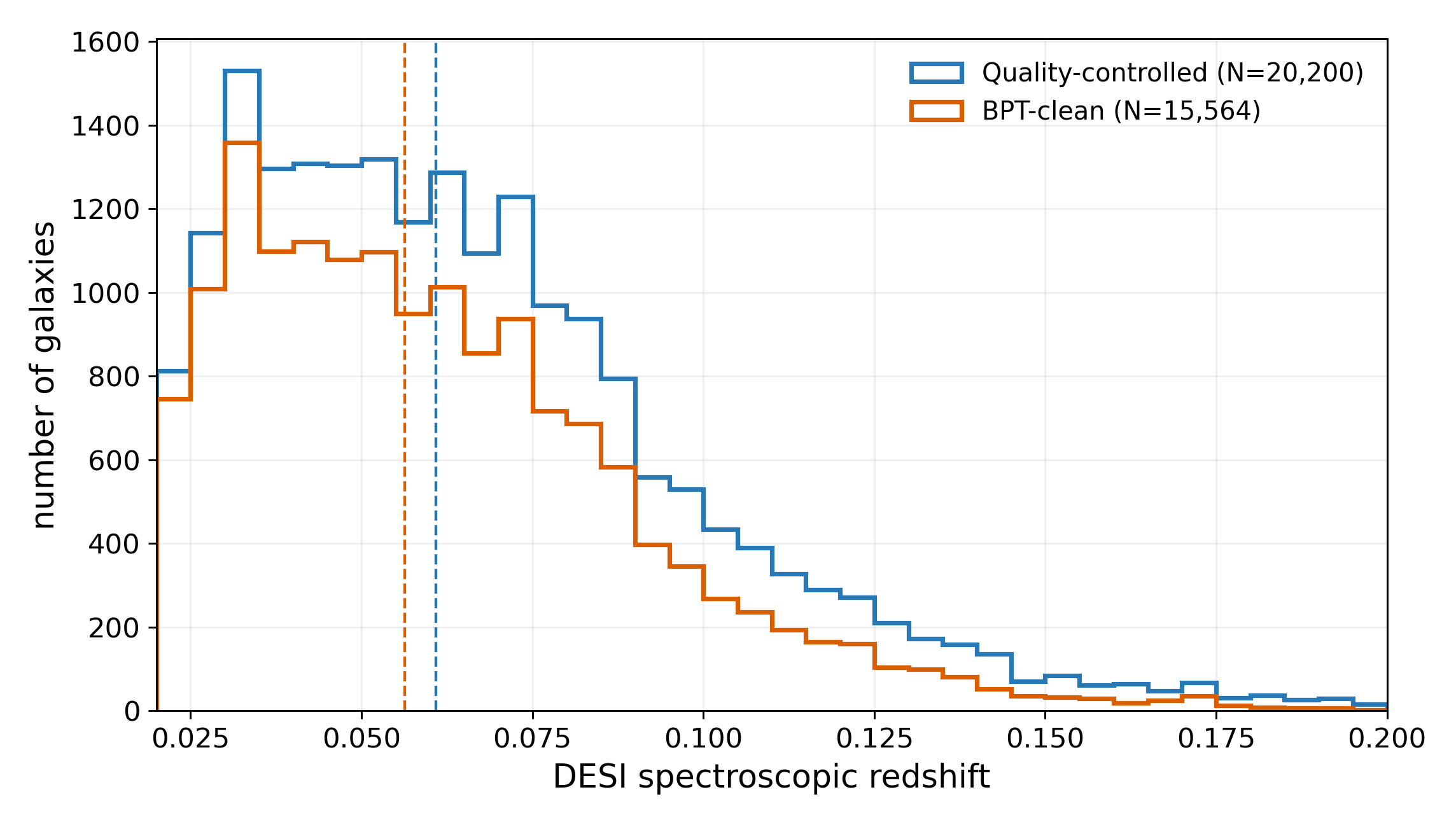}
\caption{Redshift distributions of the quality-controlled and BPT-clean samples. Dashed vertical lines mark their median DESI DR1 spectroscopic redshifts. The two samples have closely similar median redshifts, approximately 0.06.}
\label{fig:redshift_distribution}
\end{figure*}

For rest-frame fitting, we use the official DESI DR1 spectroscopic redshift for each DESI spectrum and the SDSS DR7 spectroscopic redshift for its paired SDSS spectrum. We do not determine redshifts independently. The $|\Delta v|$ requirement above ensures that using the survey-native redshift for each spectrum cannot materially alter the same-galaxy line-ratio comparison. The quality-controlled sample spans $0.020<z_{\rm DESI}<0.200$, with median $z_{\rm DESI}=0.0609$ and central 68\% interval $0.0343<z_{\rm DESI}<0.0968$; the BPT-clean subset has a closely similar median, $z_{\rm DESI}=0.0563$. Thus, both samples have median redshifts of approximately 0.06, showing that the BPT selection does not materially change the central redshift distribution. Figure~\ref{fig:redshift_distribution} shows both distributions.

\subsection{Same-pipeline spectral fitting}
\label{sec:fitmethod}
\begin{figure*}
\centering
\includegraphics[width=0.98\textwidth,height=0.78\textheight,keepaspectratio]{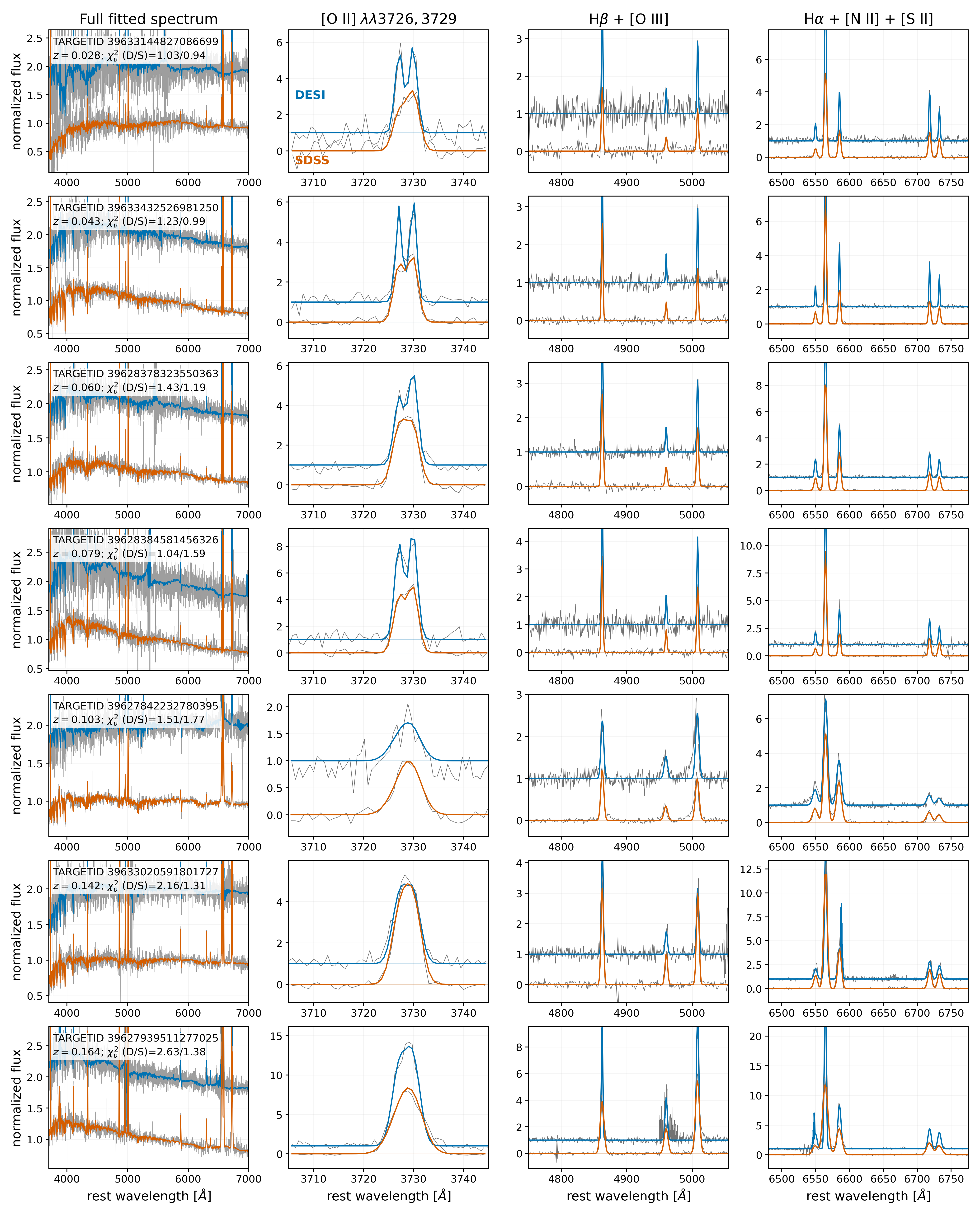}
\caption{Examples of the masked-continuum common fits. Each row shows one matched galaxy; DESI and SDSS are displayed in the same panels with separate continuum normalizations and vertical offsets. Grey curves show the spectra after foreground-extinction correction, and blue and orange curves show the corresponding total models. The full-spectrum panels include the masked-line eMILES plus residual-baseline continuum models, while the local panels show the continuum-subtracted data and gas models.}
\label{fig:samepipe_examples}
\end{figure*}

The key step is to measure DESI and SDSS spectra with the same model. We first use the DESI \texttt{FIBERMAP} colour excess $E(B-V)$ \citep{Schlegel1998,DESI2026} for both spectra in each matched pair and correct their observed-frame fluxes and inverse variances for Milky Way foreground extinction using the \citet{ODonnell1994} extinction curve with $R_V\equiv A_V/E(B-V)=3.1$, where $A_V$ is the $V$-band extinction. We do not rescale either spectrum to fibre photometry, because the difference in fibre aperture is precisely the signal under study.

We fit the stellar continuum in the observed frame over rest-frame 3600--9300~\AA\ using a non-negative linear combination of 45 single-stellar-population (SSP) templates from the eMILES library \citep{Vazdekis2016}, supplemented by an eighth-order additive Legendre polynomial. Specifically, we use the eMILES compilation distributed with pPXF \citep{Cappellari2017}, based on BaSTI isochrones \citep{Pietrinferni2004,Pietrinferni2006} and a Kroupa-like bimodal initial mass function \citep{Kroupa2001}. The basis is the Cartesian product of nine SSP-equivalent ages (0.063, 0.126, 0.251, 0.501, 1.0, 2.0, 5.01, 10.0, and 12.59~Gyr) and five total metallicities ($[\mathrm{M/H}]=-1.31$, $-0.71$, $-0.40$, 0.00, and $+0.22$). The resulting $9\times5$ basis spans the young, intermediate-age, and old populations, as well as the metal-poor to supersolar regimes required to model the stellar absorption spectra of this emission-line sample. We deliberately adopt this broad but sparse grid rather than the full set of closely spaced SSPs in the library: the continuum model serves primarily as a nuisance model for reliably removing stellar absorption, not as a measurement of a non-parametric star-formation history, and a more redundant basis would make individual template weights more sensitive to noise and degeneracies without improving the emission-line measurements. Each SSP is normalized by its median flux between 5000 and 6000~\AA\ prior to fitting.

All listed optical emission-line regions are masked by $\pm15$~\AA\ in the rest frame during the continuum fit. After subtracting the masked-line stellar model, we estimate any remaining broad baseline structure independently in each spectral segment using a 250~\AA\ observed-frame rolling median evaluated only on unmasked pixels. This is adapted from the MPA--JHU spectral-measurement procedure introduced for SDSS galaxy spectroscopy by \citet{Brinchmann2004} and \citet{Tremonti2004}: it fits a stellar model, removes low-frequency residual continuum structure, and subsequently fits emission lines with tied kinematics. Finally, we fit all optical emission lines simultaneously after removal of both the stellar and residual continua. Balmer lines share one line-of-sight velocity and velocity dispersion, while all forbidden lines, including \oii, \oiii, \nii, and \sii, share another. We fix the atomic branching ratios $[{\rm O\,III}]\lambda5007/\lambda4959=2.98$ and $[{\rm N\,II}]\lambda6583/\lambda6548=2.96$. Because the two surveys do not have identical spectral resolution, the templates and line models preserve the wavelength-dependent LSF of every individual spectrum: DESI models are convolved through the coadd resolution matrix, and SDSS models are convolved through the per-pixel \texttt{wdisp} LSF profile. The intrinsic eMILES FWHM is 2.51~\AA\ over the blue and optical MILES range and becomes wavelength dependent at the red end; at each wavelength, we broaden each stellar template only by the additional amount required to match the individual DESI or SDSS LSF. Thus, the visibly different line widths in some DESI and SDSS example spectra are explicitly modelled rather than interpreted as an aperture signal. The final nebular fluxes are therefore measured from the independent simultaneous gas-line fit after continuum subtraction, rather than inferred from the stellar-template weights.

The full fitting run returns measurements for all 20,545 matched spectra. Forty-four objects fail because one of the spectra has no usable fitted segment over the adopted rest-frame range. We quantify fit quality using the standard reduced chi-square,
\begin{equation}
\chi^2_\nu=\frac{\sum_i[(F_i-F_{{\rm model},i})/\sigma_i]^2}{N_{\rm pix}-N_{\rm par}},
\end{equation}
where $F_i$, $F_{{\rm model},i}$, and $\sigma_i$ are the flux, model flux, and 1$\sigma$ flux uncertainty of valid pixel $i$, respectively; $N_{\rm pix}$ is the number of valid pixels; and $N_{\rm par}$ is the number of fitted parameters. We retain successful pairs with $\chi^2_\nu<3$ in both surveys, yielding a fiducial sample of 20,200 galaxies.

Before interpreting aperture offsets, we verify that the common refit provides sensible line measurements for both surveys. Figure~\ref{fig:samepipe_examples} shows representative DESI--SDSS spectral pairs. The full-spectrum panels assess the stellar-continuum fit, while the local panels assess the line complexes used in the aperture analysis. The two spectra are not expected to have identical fluxes because their fibre apertures differ. Nevertheless, the common continuum and emission-line methodology provides a physically sensible description of both surveys.

\section{Results}
\label{sec:fitquality}

\subsection{Aperture Offsets in Line Ratios}
\label{sec:results}
For each galaxy, we compute the same line ratio from the DESI and SDSS refit fluxes and define $\Delta X=X_{\rm DESI}-X_{\rm SDSS}$. We report median offsets and estimate their uncertainties by bootstrap resampling of galaxies \citep{Efron1979}.

In addition to the 20,200 quality-controlled same-pipeline sample, we define a BPT-clean star-forming subset by requiring both DESI and SDSS measurements to lie below the \citet{Kauffmann2003} star-forming demarcation in the \nii\ BPT diagram; this leaves 15,564 pairs. We also define a near-concentric subset by further requiring a fibre-centre separation $<0.10$ arcsec and $|\Delta v|<20~\kms$, leaving 4,607 BPT-clean pairs.

\begin{figure*}
\centering
\includegraphics[width=0.95\textwidth]{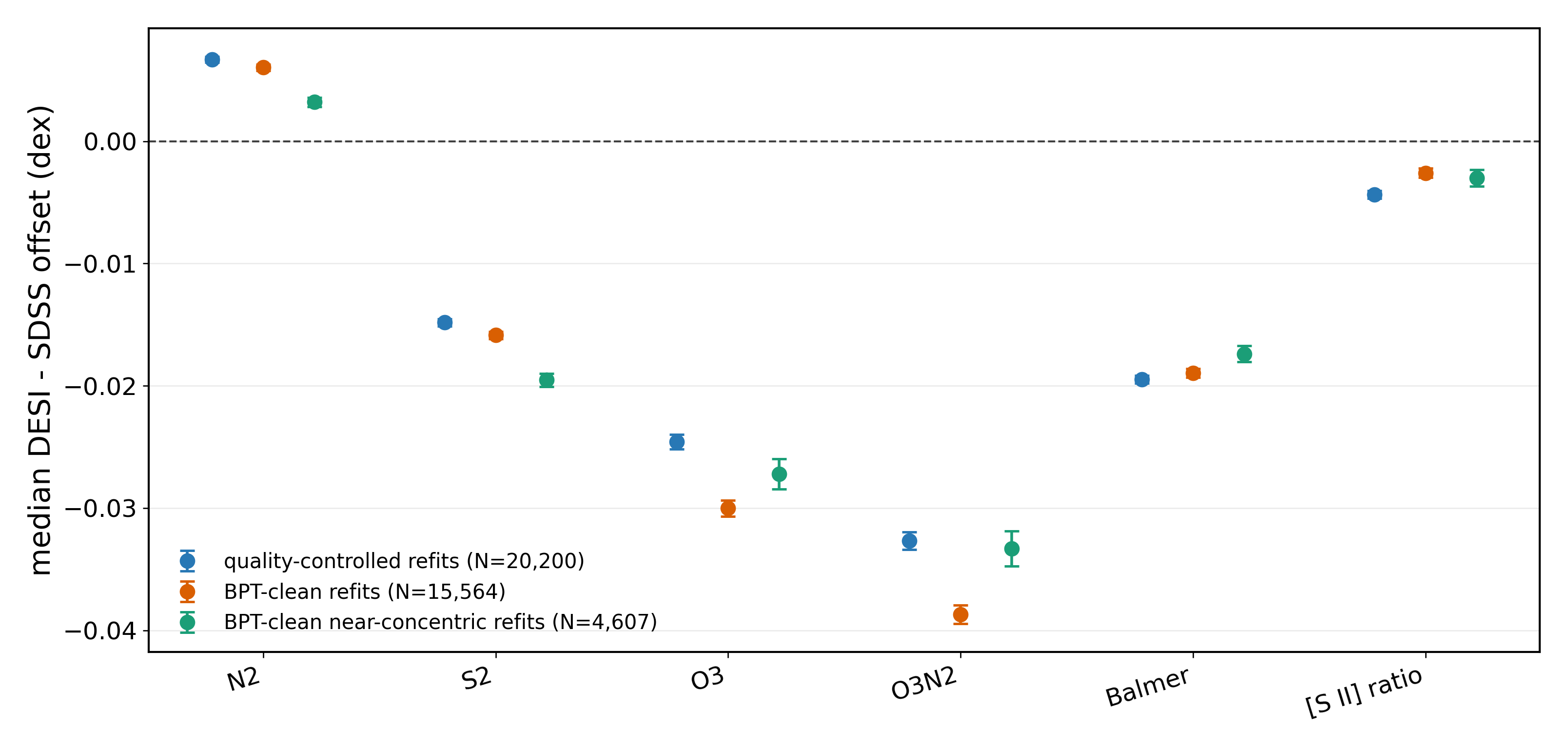}
\caption{Median DESI--SDSS line-ratio offsets from the same-pipeline spectral refit. Blue points show the quality-controlled pairs with $\chi^2_\nu<3$ in both surveys, orange points show the BPT-clean star-forming subset, and green points show the BPT-clean near-concentric subset with separation $<0.10$ arcsec and $|\Delta v|<20~\kms$. Error bars show bootstrap uncertainties on the median. The robust aperture terms are negative offsets in S2, O3, O3N2, and the Balmer decrement; the N2 offset is positive but weaker.}
\label{fig:offsets}
\end{figure*}

Figure~\ref{fig:offsets} presents the main aperture result. In the 20,200 quality-controlled same-pipeline pairs, the median offsets are
\begin{align}
\Delta\log{\rm N2} &= +0.0067\pm0.0002,\\
\Delta\log{\rm S2} &= -0.0148\pm0.0003,\\
\Delta\log{\rm O3} &= -0.0246\pm0.0006,\\
\Delta{\rm O3N2} &= -0.0327\pm0.0007,\\
\Delta\log({\rm H}\alpha/{\rm H}\beta) &= -0.0195\pm0.0003,\\
\Delta\log([{\rm S\,II}]6716/6731) &= -0.0044\pm0.0003 .
\end{align}

The BPT-clean subset shows the same qualitative pattern:
\begin{align}
\Delta\log{\rm N2} &=+0.0060\pm0.0003,\\
\Delta\log{\rm S2} &=-0.0159\pm0.0003,\\
\Delta\log{\rm O3} &=-0.0300\pm0.0007,\\
\Delta{\rm O3N2} &=-0.0387\pm0.0007, \\
\Delta\log({\rm H}\alpha/{\rm H}\beta) &=-0.0190\pm0.0004,\\
\Delta\log([{\rm S\,II}]6716/6731) &= -0.0026\pm0.0004 .
\end{align}
Thus, the result is not driven by obvious AGN or low-ionization nuclear emission-line region contamination.

The BPT-clean near-concentric subset yields
\begin{align}
\Delta\log{\rm N2} &=+0.0032\pm0.0004,\\
\Delta\log{\rm S2} &=-0.0195\pm0.0005,\\
\Delta\log{\rm O3} &=-0.0272\pm0.0012,\\
\Delta{\rm O3N2} &=-0.0333\pm0.0014, \\
\Delta\log({\rm H}\alpha/{\rm H}\beta) &=-0.0174\pm0.0007,\\
\Delta\log([{\rm S\,II}]6716/6731) &= -0.0030\pm0.0007 .
\end{align}
This strict-centering result is important: the N2 aperture term is small and sensitive to centering, whereas the S2, O3, O3N2, and Balmer-decrement terms remain robust.

\begin{figure*}
\centering
\includegraphics[width=0.98\textwidth]{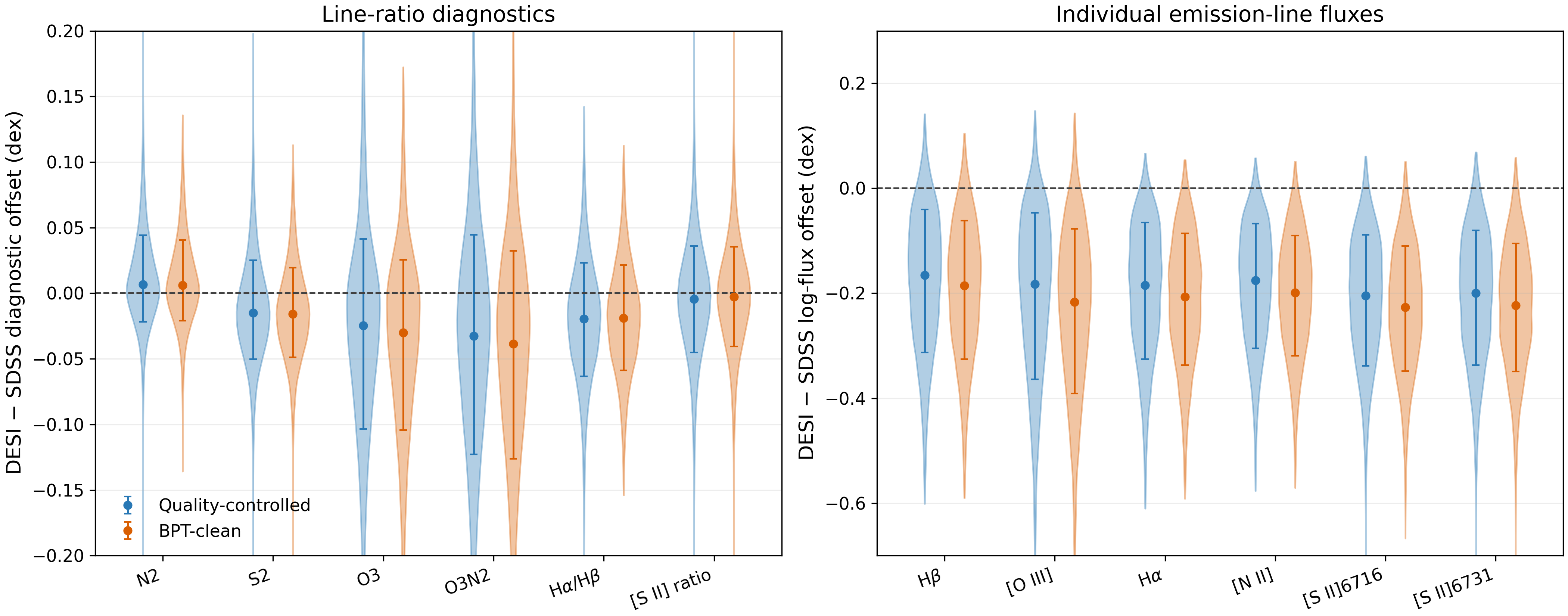}
\caption{Full distributions of same-pipeline DESI--SDSS differences. Left: the six diagnostic offsets. Right: individual-line log-flux offsets. Violin widths trace the distribution after clipping only the most extreme 0.5\% tails for display; coloured points and bars give the median and central 16th--84th percentile interval of the untrimmed distributions. Blue and orange denote the quality-controlled and BPT-clean samples. The broad, non-Gaussian distributions show why the median describes a net population-level aperture term rather than an object-by-object correction.}
\label{fig:offset_distributions}
\end{figure*}

The median in Figure~\ref{fig:offsets} summarizes a net population-level term and does not imply Gaussian or object-by-object identical differences. Figure~\ref{fig:offset_distributions} shows the full per-galaxy distributions of the six diagnostic offsets and the corresponding individual-line flux offsets. The distributions are broad and have asymmetric tails: for the quality-controlled sample, their central 16th--84th percentile ranges span $-0.022$ to $+0.044$ dex for N2, $-0.050$ to $+0.025$ dex for S2, $-0.104$ to $+0.041$ dex for O3, and $-0.123$ to $+0.044$ dex for O3N2. The BPT-clean distributions are similarly broad. These widths combine genuine galaxy-to-galaxy aperture variation with measurement noise, centering residuals, seeing, and other survey-to-survey effects. The error bars on the medians indicate how precisely the median offset is determined for the sample; they do not represent the range of offsets expected for individual galaxies.

\begin{figure*}
\centering
\includegraphics[width=0.82\textwidth]{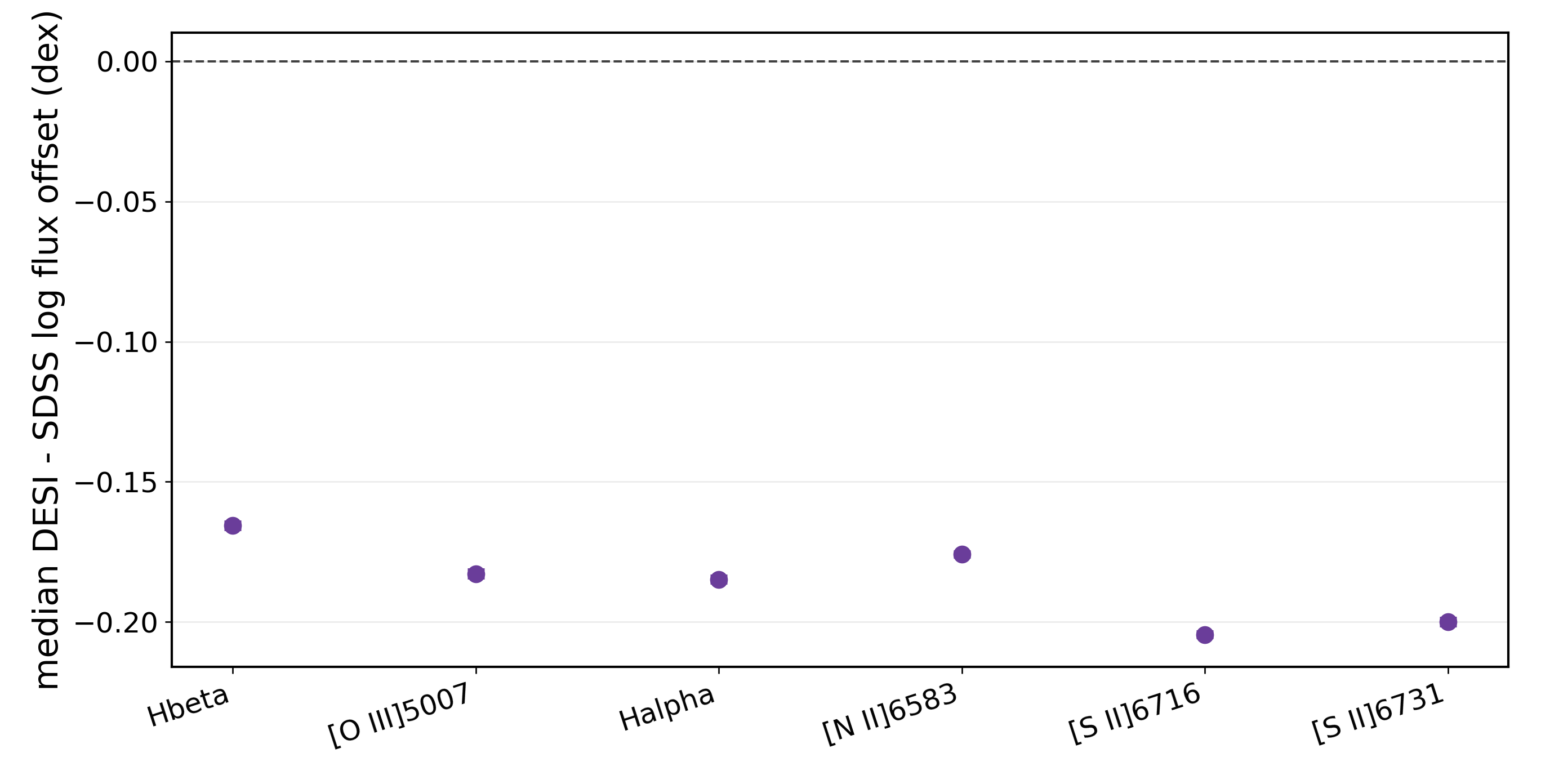}
\caption{Individual-line DESI--SDSS flux offsets from the successful masked-continuum refits. The useful information in this figure is not simply that all DESI fluxes are lower than their SDSS counterparts, which is expected because the DESI fibre is smaller. The key point is that the offsets are not identical for all lines. A single grey aperture scaling would place all six points at the same value and would leave all line ratios unchanged. 
}
\label{fig:fluxloss}
\end{figure*}

The individual emission-line fluxes themselves are not the main result, but they clarify how the line-ratio offsets arise. As expected for nested apertures, DESI line fluxes are lower than SDSS line fluxes for all six diagnostic lines. If the smaller DESI aperture simply multiplied every emission line by the same factor, all points in Fig.~\ref{fig:fluxloss} would show the same vertical offset, and line ratios such as N2, S2, O3, O3N2, and H$\alpha$/H$\beta$ would remain unchanged. Instead, the median DESI--SDSS log-flux offset is $-0.16$ dex for \hb, $-0.18$ dex for \ha, and $-0.20$ dex for \sii$\lambda6716$ and \sii$\lambda6731$, respectively. The aperture loss is therefore line dependent rather than a simple grey scaling, which naturally gives rise to the line-ratio offsets measured above.

\section{Discussion}
\label{sec:discussion}

\subsection{Plausible physical interpretation}
The same-pipeline offsets are modest in amplitude but coherent in sign and significance. The smaller DESI fibre loses different fractions of different line-emitting components depending on their spatial distributions. Spatially resolved surveys have established that nebular line ratios vary systematically within galaxies and that diffuse ionized gas can materially alter low-ionization line ratios \citep{Belfiore2017,Zhang2017}. The flux pattern is therefore consistent with, but does not uniquely establish, a contribution from more extended lower-ionization \sii-emitting gas or from diffuse ionized gas sampled preferentially by the larger SDSS fibre. Because metallicity, ionization parameter, dust, and diffuse-gas fraction can covary with radius, and because this work does not derive these quantities spatially for every galaxy, we treat this as a qualitative physical interpretation rather than a line-by-line causal decomposition. The negative O3 and O3N2 offsets imply that the aperture effect is not simply a matter of enhanced central excitation in this high-S/N sample; instead, the relative spatial distributions of \oiii, \hb, \nii, and \ha\ combine to produce a net lower DESI O3N2. The lower DESI Balmer decrement is recovered in all clean and BPT-clean tests and may reflect aperture-dependent dust geometry, residual stellar-absorption systematics, or different spatial distributions of Balmer-line emission.

The existence of aperture-dependent nebular diagnostics is well established, although the detailed growth curves depend on galaxy population and on the apertures being compared. Nuclear-to-integrated comparisons have found aperture-dependent reddening and strong-line abundance indicators \citep{Kewley2005,Moustakas2006}. Using CALIFA aperture growth curves, \citet{IglesiasParamo2016} further showed that the Balmer decrement, N2, and O3N2 all vary systematically with enclosed aperture. Related targeted work on the nuclear spectrum of UGC~2885 illustrates that observed central line ratios can differ from other aperture-matched spectroscopic measurements of the same galaxy \citep{Holwerda2021}. Our results extend this empirical picture to the direct DESI--SDSS overlap: with a common continuum, line, and LSF treatment, we detect diagnostic-dependent offsets between nested 1.5- and 3-arcsec fibres. We do not interpret the present measurements as a universal aperture correction, because neither fibre provides an integrated measurement and because the amplitude, and even the sign, of an aperture term can depend on the radial line-ratio structure, sample selection, seeing, and fibre centering.

The weak positive N2 offset is more sensitive to centering. In the full and BPT-clean samples it is statistically significant, but its amplitude decreases in the BPT-clean near-concentric subset. We therefore do not regard N2 alone as the most robust aperture-bias diagnostic. The stronger conclusion is that DESI--SDSS comparisons exhibit a multi-line aperture term at the $0.015$--$0.03$ dex level in S2, O3, O3N2, and the Balmer decrement, with a smaller N2 component whose amplitude depends on centering and sample selection.

\subsection{Why the effect matters}
The offsets measured here are much smaller than the absolute discrepancies among different strong-line abundance calibrations, but they are comparable to the precision targeted by modern differential studies. A shift of $0.01$--$0.03$ dex in O3N2 or in the Balmer decrement can matter when combining DESI and SDSS samples, measuring weak environmental trends, or recalibrating line-ratio relations using mixed-aperture data. The safest practice is therefore to treat DESI and SDSS line ratios as aperture-dependent measurements unless an aperture correction or matched-aperture modelling step is applied.

\subsection{Redshift and angular-size dependence}
\label{sec:redshift_size}
\begin{figure*}
\centering
\includegraphics[width=0.98\textwidth]{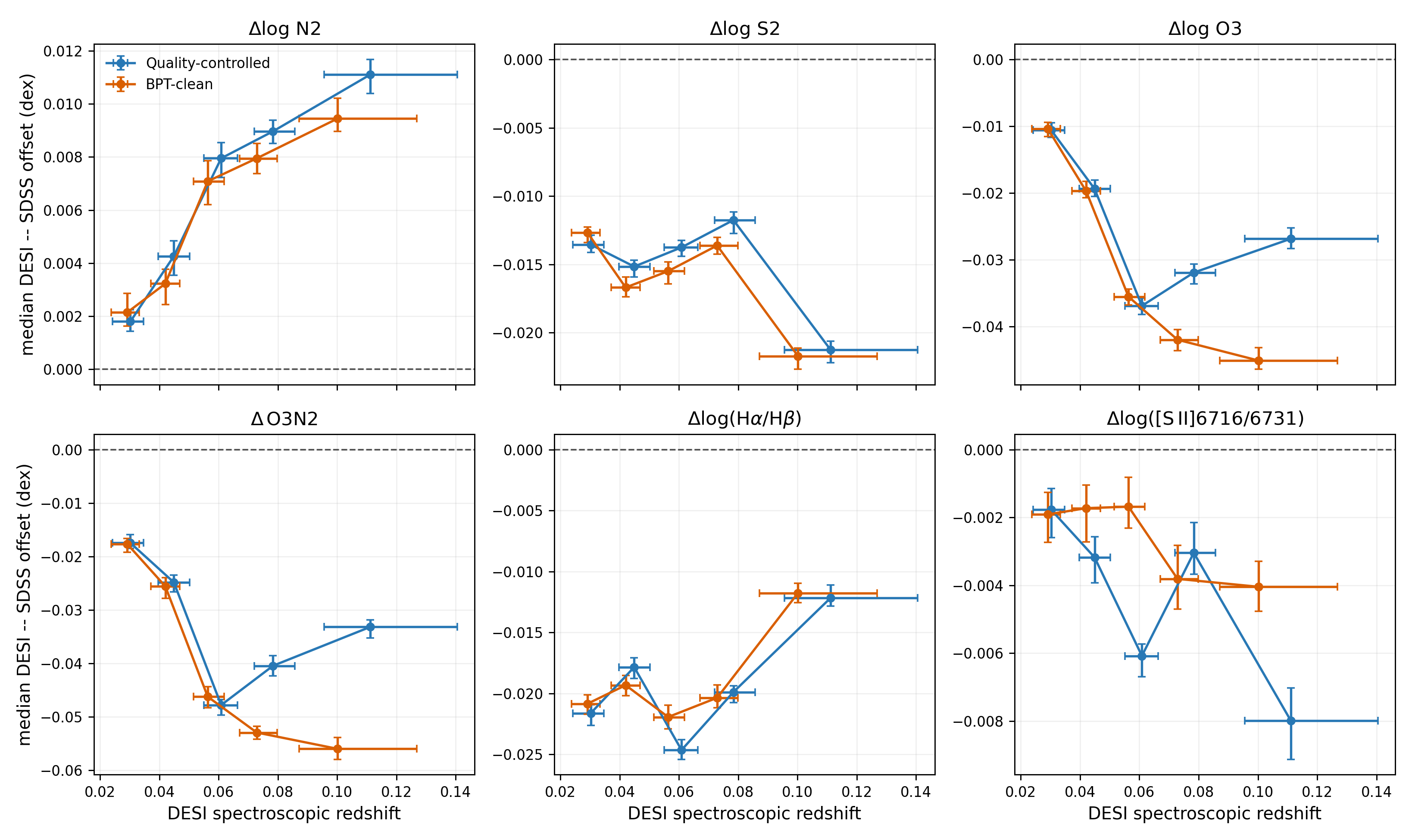}
\caption{Redshift dependence of all six same-pipeline DESI--SDSS diagnostic offsets. Points show equal-number redshift bins; horizontal bars give their central 68\% redshift ranges and vertical bars are bootstrap uncertainties on the median. Blue and orange curves show the quality-controlled and BPT-clean samples, respectively. The aperture terms are not constant with redshift, as expected because a fixed angular fibre samples a changing physical and relative galactic scale.}
\label{fig:redshift_dependence}
\end{figure*}

\begin{figure*}
\centering
\includegraphics[width=0.98\textwidth]{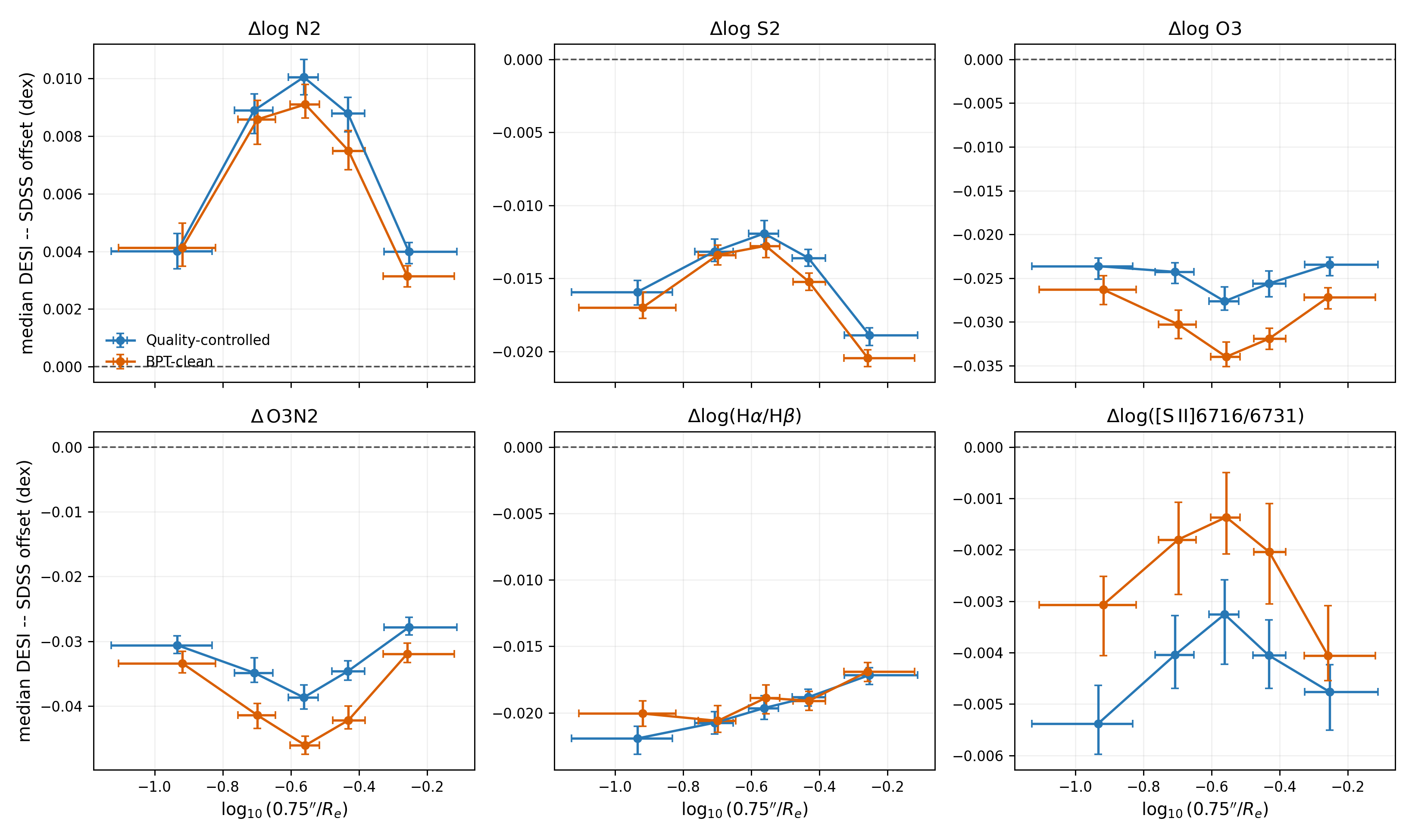}
\caption{Dependence of the six diagnostic offsets on apparent fibre coverage. $R_e$ is the Legacy Surveys major-axis angular effective (half-light) radius, measured in arcsec and reported in the DESI DR1 value-added catalogue; thus $0.75^{\prime\prime}/R_e$ is dimensionless, and a larger value means that the DESI fibre samples a larger fraction of the apparent galaxy. Points show equal-number bins and bootstrap uncertainties on the median.}
\label{fig:coverage_dependence}
\end{figure*}

The fixed fibre diameters imply that both physical aperture scale and relative aperture coverage change across the sample. We therefore repeat the same-pipeline measurement in five equal-number bins of DESI DR1 redshift and, separately, in five bins of the dimensionless apparent-coverage proxy $0.75^{\prime\prime}/R_e$, where $R_e$ is the catalogue major-axis angular effective (half-light) radius in arcsec. Figures~\ref{fig:redshift_dependence} and \ref{fig:coverage_dependence} give the results for all six diagnostics and for both the quality-controlled and BPT-clean samples.

The offsets are not redshift invariant. In the quality-controlled sample, the median O3 offset changes from $-0.0106$ dex at $z=0.030$ to $-0.0268$ dex at $z=0.111$, and O3N2 changes from $-0.0175$ to $-0.0332$ dex over the same bins. These changes are larger than the bootstrap uncertainties of the corresponding bin medians and are clearest for O3 and O3N2. N2 rises from $+0.0018$ to $+0.0111$ dex, while S2 remains negative in every bin ($-0.0118$ to $-0.0213$ dex); the Balmer-decrement and \sii-ratio offsets also remain predominantly negative, although their amplitudes vary non-monotonically. The BPT-clean sample follows the same qualitative behavior. We interpret these changes as evidence that a fixed angular fibre samples different physical and relative galactic scales, combined with radial variation in the spatial distributions of the line-emitting components. They should not be interpreted as a measurement of intrinsic redshift evolution, because redshift is also correlated with the galaxy population and with the selection function. Thus, a single full-sample offset is an informative summary for the selected redshift distribution, but not a redshift-independent aperture correction.

As a complementary check, we measure the quality-controlled sample in three fixed redshift windows centred near $z=0.05$, 0.10, and 0.15: $0.04<z<0.06$ ($N=5100$), $0.09<z<0.11$ ($N=1911$), and $0.14<z<0.16$ ($N=351$). The offsets remain diagnostic dependent in all three windows. For example, $\Delta\log{\rm O3}=-0.0240\pm0.0010$, $-0.0260\pm0.0018$, and $-0.0401^{+0.0082}_{-0.0105}$ dex, respectively; the corresponding O3N2 offsets are $-0.0318\pm0.0017$, $-0.0334\pm0.0025$, and $-0.0433^{+0.0103}_{-0.0141}$ dex. The highest-redshift window is necessarily less precise because it contains fewer paired high-S/N spectra.

\begin{figure*}
\centering
\includegraphics[width=0.98\textwidth]{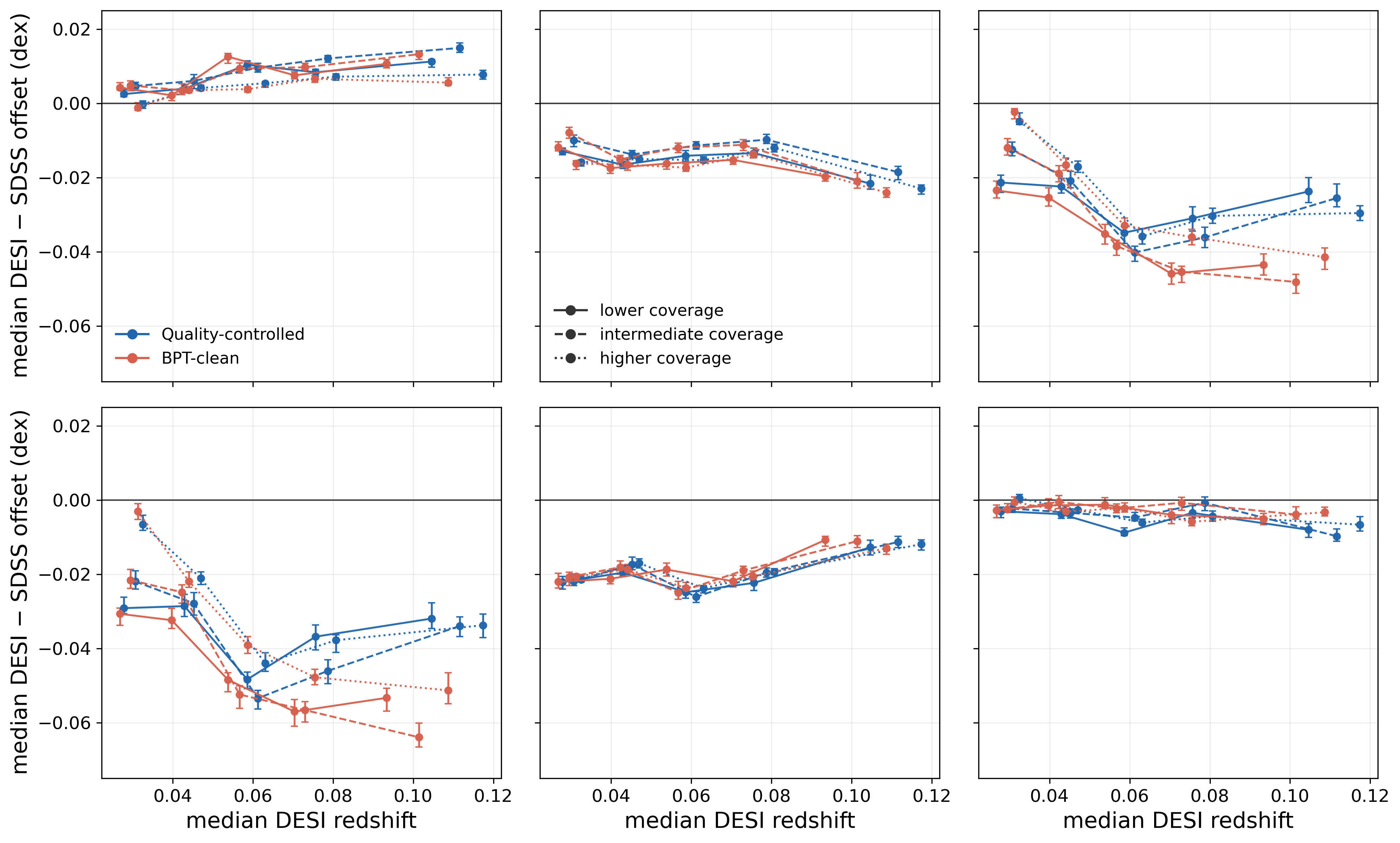}
\caption{Absolute DESI--SDSS offsets in three apparent-coverage tertiles within each of five equal-number redshift bins. Blue and red symbols denote the quality-controlled and BPT-clean samples, respectively; solid, dashed, and dotted lines denote lower, intermediate, and higher apparent coverage. The principal aperture-sensitive diagnostics remain non-zero across the coverage tertiles, and the systematic separation between coverage groups is most evident for O3 and O3N2.}
\label{fig:redshift_controlled_coverage}
\end{figure*}

The apparent-coverage trends in Figure~\ref{fig:coverage_dependence} remain when redshift is controlled. We divide each sample into five equal-number redshift bins and, within each redshift bin, compare the lower, intermediate, and higher tertiles of the dimensionless apparent-coverage proxy $0.75^{\prime\prime}/R_e$. Figure~\ref{fig:redshift_controlled_coverage} shows the resulting absolute offsets. The principal aperture-sensitive diagnostics remain systematically offset from zero across the coverage tertiles, while their amplitudes change with apparent fibre coverage. This coverage-dependent separation is especially clear for O3 and O3N2: the weighted higher-minus-lower coverage differences are $+0.0031$ and $+0.0063$ dex in the quality-controlled sample and $+0.0088$ and $+0.0117$ dex in the BPT-clean sample, respectively. Thus, the DESI--SDSS aperture bias is not only redshift dependent; its amplitude also changes with the fraction of the apparent galaxy enclosed by the DESI fibre. This result supports the interpretation that radial line-ratio structure, sampled at different relative aperture sizes, jointly shapes the diagnostic-dependent offsets.

\subsection{Scope and Limitations}
The main limitation of this study is its scope. Requiring high S/N in the six diagnostic emission lines selects emission-line-bright systems and underrepresents dusty, weak-line, and quiescent galaxies. Accordingly, the quantitative offsets reported here apply specifically to high-S/N emission-line galaxies and should not be treated as universal corrections for the full DESI--SDSS galaxy overlap. The aperture behaviour of weak-emission-line systems may differ and requires a dedicated analysis. Our common analysis uses eMILES rather than the distinct stellar libraries adopted in the public DESI and SDSS catalogues; the masked-line continuum fit and residual-baseline correction reduce, but do not eliminate, stellar-model systematics. Finally, this work measures empirical fibre-aperture terms. It does not attempt a physical decomposition into metallicity gradients, dust geometry, diffuse ionized gas, and nuclear excitation, which would require imaging data and spatially resolved spectroscopy. The broad tails in Figure~\ref{fig:offset_distributions} also mean that rare pathological pairs should not be interpreted through the median result. Blended or overlapping galaxy pairs and strong gravitational lenses can produce unusual composite spectra \citep{Holwerda2015,Holwerda2022,Bolton2006}; identifying such systems object by object requires imaging and lens-selection information beyond the present catalogue-level experiment. As demonstrated in Section~\ref{sec:redshift_size}, even within a homogeneous fibre survey the sampled relative galaxy area changes with redshift; analyses seeking redshift evolution in line diagnostics should therefore model aperture coverage or compare samples with matched redshift and size distributions.

\section{Conclusions}
\label{sec:conclusions}
We have performed a direct DESI--SDSS two-aperture comparison for 20,545 galaxies with high-confidence measurements of six optical diagnostic lines in both surveys. All spectra were refitted with a common masked-continuum eMILES plus emission-line model that incorporates the wavelength-dependent spectral resolutions of both DESI and SDSS. Our main conclusions are as follows:
\begin{enumerate}
\item Of the 20,545 matched spectra, 20,501 were fitted successfully and 20,200 passed the common-fit quality criterion $\chi^2_\nu<3$ in both surveys. The remaining 44 fitting failures and 301 pairs rejected by the quality cut are excluded from the fiducial statistics.
\item The same-pipeline line-ratio offsets are small but highly significant. We find $\Delta\log{\rm N2}=+0.0067\pm0.0002$, $\Delta\log{\rm S2}=-0.0148\pm0.0003$, and $\Delta\log{\rm O3}=-0.0246\pm0.0006$. We further find $\Delta{\rm O3N2}=-0.0327\pm0.0007$ and $\Delta\log({\rm H}\alpha/{\rm H}\beta)=-0.0195\pm0.0003$.
\item In a BPT-clean star-forming subset of 15,564 galaxies, the same qualitative pattern remains: N2 is weakly positive, whereas S2, O3, O3N2, and the Balmer decrement are negative.
\item In the BPT-clean near-concentric subset, the N2 offset becomes smaller, whereas S2, O3, O3N2, and the Balmer decrement remain significantly negative. These latter diagnostics therefore provide the most robust aperture-bias signal.
\item The six offsets vary measurably across the redshift distribution and retain a measurable dependence on the apparent fibre-coverage proxy after redshift is controlled. Redshift and relative aperture coverage therefore jointly modulate the diagnostic-dependent aperture bias, so the full-sample values are not universal corrections.
\item For high-S/N emission-line galaxies, the net effect is a small but highly significant aperture term that should be taken into account in precision DESI--SDSS comparisons and in future DESI strong-line calibration work. The present results should not be interpreted as a universal correction for weak-emission-line or quiescent systems.
\end{enumerate}

\vspace{6pt}

\begin{acknowledgments}
	
Y.R. acknowledges supports from the CAS Pioneer Hundred Talents Program (Category B), and the NSFC grants 12522302, 12673017, and 12273037. This work is also supported by the China Manned Space Program with grant nos.\ CMS-CSST-2025-A06 and CMS-CSST-2025-A08. The data used in this study are publicly available from the DESI and SDSS survey archives. Additional processed data products are available from the corresponding author upon reasonable request. This work makes use of data from DESI and SDSS. We thank the DESI and SDSS collaborations for making their survey products publicly available. In this work, we use AI to select the DESI-SDSS fibre-matched sample.

\end{acknowledgments}





\begin{thebibliography}{}
\bibitem[\protect\citeauthoryear{Abazajian et al.}{2009}]{Abazajian2009} Abazajian, K. N., Adelman-McCarthy, J. K., Ag\"ueros, M. A., Allam, S. S., Allende Prieto, C., An, D., Anderson, K. S. J., Anderson, S. F., Annis, J., Bahcall, N. A., et al., ``The Seventh Data Release of the Sloan Digital Sky Survey'', 2009, ApJS, 182, 543
\bibitem[\protect\citeauthoryear{Baldwin, Phillips \& Terlevich}{1981}]{Baldwin1981} Baldwin, J. A., Phillips, M. M., Terlevich, R., ``Classification parameters for the emission-line spectra of extragalactic objects.'', 1981, PASP, 93, 5
\bibitem[\protect\citeauthoryear{Belfiore et al.}{2017}]{Belfiore2017} Belfiore, F., Maiolino, R., Maraston, C., Emsellem, E., Bershady, M. A., Masters, K. L., Bizyaev, D., Boquien, M., Brownstein, J. R., Bundy, K., et al., ``SDSS-IV MaNGA - the spatially resolved transition from star formation to quiescence'', 2017, MNRAS, 466, 2570
\bibitem[\protect\citeauthoryear{Belfiore et al.}{2017}]{Belfiore2017b} Belfiore, F., Maiolino, R., Tremonti, C., S\'anchez, S. F., Bundy, K., Bershady, M., Westfall, K., Lin, L., Drory, N., Boquien, M., et al., ``SDSS IV MaNGA - metallicity and nitrogen abundance gradients in local galaxies'', 2017, MNRAS, 469, 151
\bibitem[\protect\citeauthoryear{Bolton et al.}{2006}]{Bolton2006} Bolton, A. S., Burles, S., Koopmans, L. V. E., Treu, T., Moustakas, L. A., ``The Sloan Lens ACS Survey. I. A Large Spectroscopically Selected Sample of Massive Early-Type Lens Galaxies'', 2006, ApJ, 638, 703
\bibitem[\protect\citeauthoryear{Brinchmann et al.}{2004}]{Brinchmann2004} Brinchmann, J., Charlot, S., White, S. D. M., Tremonti, C., Kauffmann, G., Heckman, T., Brinkmann, J., ``The physical properties of star-forming galaxies in the low-redshift Universe'', 2004, MNRAS, 351, 1151
\bibitem[\protect\citeauthoryear{Cappellari}{2017}]{Cappellari2017} Cappellari, M., ``Improving the full spectrum fitting method: accurate convolution with Gauss-Hermite functions'', 2017, MNRAS, 466, 798
\bibitem[\protect\citeauthoryear{Bundy et al.}{2015}]{Bundy2015} Bundy, K., Bershady, M. A., Law, D. R., Yan, R., Drory, N., MacDonald, N., Wake, D. A., Cherinka, B., S\'anchez-Gallego, J. R., Weijmans, A.-M., et al., ``Overview of the SDSS-IV MaNGA Survey: Mapping nearby Galaxies at Apache Point Observatory'', 2015, ApJ, 798, 7
\bibitem[\protect\citeauthoryear{Croom et al.}{2012}]{Croom2012} Croom, S. M., Lawrence, J. S., Bland-Hawthorn, J., Bryant, J. J., Fogarty, L., Richards, S., Goodwin, M., Farrell, T., Miziarski, S., Heald, R., et al., ``The Sydney-AAO Multi-object Integral field spectrograph'', 2012, MNRAS, 421, 872
\bibitem[\protect\citeauthoryear{Curti et al.}{2017}]{Curti2017} Curti, M., Cresci, G., Mannucci, F., Marconi, A., Maiolino, R., Esposito, S., ``New fully empirical calibrations of strong-line metallicity indicators in star-forming galaxies'', 2017, MNRAS, 465, 1384
\bibitem[\protect\citeauthoryear{Curti et al.}{2020}]{Curti2020} Curti, M., Mannucci, F., Cresci, G., Maiolino, R., ``The mass-metallicity and the fundamental metallicity relation revisited on a fully $T_e$-based abundance scale for galaxies'', 2020, MNRAS, 491, 944
\bibitem[\protect\citeauthoryear{DESI Collaboration et al.}{2016}]{Aghamousa2016} DESI Collaboration, Aghamousa, A., Aguilar, J., Ahlen, S., Alam, S., Allen, L. E., Allende Prieto, C., Annis, J., Bailey, S., Balland, C., et al., ``The DESI Experiment Part I: Science,Targeting, and Survey Design'', 2016, arXiv:1611.00036
\bibitem[\protect\citeauthoryear{DESI Collaboration et al.}{2022}]{DESI2022} DESI Collaboration, Abareshi, B., Aguilar, J., Ahlen, S., Alam, S., Alexander, D. M., Alfarsy, R., Allen, L., Allende Prieto, C., Alves, O., et al., ``Overview of the Instrumentation for the Dark Energy Spectroscopic Instrument'', 2022, AJ, 164, 207
\bibitem[\protect\citeauthoryear{DESI Collaboration et al.}{2026}]{DESI2026} DESI Collaboration, Abdul Karim, M., Adame, A. G., Aguado, D., Aguilar, J., Ahlen, S., Alam, S., Aldering, G., Alexander, D. M., Alfarsy, R., et al., ``Data Release 1 of the Dark Energy Spectroscopic Instrument'', 2026, AJ, 171, 285
\bibitem[\protect\citeauthoryear{Efron}{1979}]{Efron1979} Efron, B., ``Bootstrap Methods: Another Look at the Jackknife'', 1979, Ann. Statist., 7, 1
\bibitem[\protect\citeauthoryear{Ellison et al.}{2009}]{Ellison2009} Ellison, S. L., Simard, L., Cowan, N. B., Baldry, I. K., Patton, D. R., McConnachie, A. W., ``The mass-metallicity relation in galaxy clusters: the relative importance of cluster membership versus local environment'', 2009, MNRAS, 396, 1257
\bibitem[\protect\citeauthoryear{Hopkins et al.}{2003}]{Hopkins2003} Hopkins, A. M., Miller, C. J., Nichol, R. C., Connolly, A. J., Bernardi, M., G\'omez, P. L., Goto, T., Tremonti, C. A., Brinkmann, J., Ivezi\'c, \v{Z}., Lamb, D. Q., ``Star Formation Rate Indicators in the Sloan Digital Sky Survey'', 2003, ApJ, 599, 971
\bibitem[\protect\citeauthoryear{Holwerda et al.}{2015}]{Holwerda2015} Holwerda, B. W., Baldry, I. K., Alpaslan, M., Bauer, A., Bland-Hawthorn, J., Brough, S., Brown, M. J. I., Cluver, M. E., Conselice, C., Driver, S. P., et al., ``Galaxy And Mass Assembly (GAMA) blended spectra catalogue: strong galaxy-galaxy lens and occulting galaxy pair candidates'', 2015, MNRAS, 449, 4277
\bibitem[\protect\citeauthoryear{Holwerda et al.}{2021}]{Holwerda2021} Holwerda, B. W., Wu J. F., Keel W. C., Young, J., Mullins, R., Hinz, J., Ford, K. E. Saavik, Barmby, P., Chandar, R., Bailin, J., et al., ``Predicting the Spectrum of UGC 2885, Rubin's Galaxy with Machine Learning'', 2021, ApJ, 914, 142
\bibitem[\protect\citeauthoryear{Holwerda et al.}{2022}]{Holwerda2022} Holwerda, B. W., Knabel, S., Thorne, J. E., Bellstedt, S., Siudek, M., Davies, L. J. M., ``Deep Extragalactic VIsible Legacy Survey: Data Release 1 blended spectra search for candidate strong gravitational lenses'', 2022, MNRAS, 510, 2305
\bibitem[\protect\citeauthoryear{Iglesias-P{\'a}ramo et al.}{2016}]{IglesiasParamo2016} Iglesias-P{\'a}ramo, J., V\'ilchez, J. M., Rosales-Ortega, F. F., S\'anchez, S. F., Duarte Puertas, S., Petropoulou, V., Gil de Paz, A., Galbany, L., Moll\'a, M., Catal\'an-Torrecilla, C., et al., ``Aperture Effects on the Oxygen Abundance Determinations from CALIFA Data'', 2016, ApJ, 826, 71
\bibitem[\protect\citeauthoryear{Kauffmann et al.}{2003}]{Kauffmann2003} Kauffmann, G., Heckman, T. M., Tremonti, C., Brinchmann, J., Charlot, S., White, S. D. M., Ridgway, S. E., Brinkmann, J., Fukugita, M., Hall, P. B., et al., ``The host galaxies of active galactic nuclei'', 2003, MNRAS, 346, 1055
\bibitem[\protect\citeauthoryear{Kewley et al.}{2001}]{Kewley2001} Kewley, L. J., Dopita, M. A., Sutherland, R. S., Heisler, C. A., Trevena, J., ``Theoretical Modeling of Starburst Galaxies'', 2001, ApJ, 556, 121
\bibitem[\protect\citeauthoryear{Kewley, Jansen \& Geller}{2005}]{Kewley2005} Kewley, L. J., Jansen, R. A., Geller, M. J., ``Aperture Effects on Star Formation Rate, Metallicity, and Reddening'', 2005, PASP, 117, 227
\bibitem[\protect\citeauthoryear{Kewley \& Ellison}{2008}]{Kewley2008} Kewley, L. J., Ellison S. L., ``Metallicity Calibrations and the Mass-Metallicity Relation for Star-forming Galaxies'', 2008, ApJ, 681, 1183
\bibitem[\protect\citeauthoryear{Kewley, Nicholls \& Sutherland}{2019}]{Kewley2019} Kewley, L. J., Nicholls, D. C., Sutherland, R. S., ``Understanding Galaxy Evolution Through Emission Lines'', 2019, ARA\&A, 57, 511
\bibitem[\protect\citeauthoryear{Kroupa}{2001}]{Kroupa2001} Kroupa, P., ``On the variation of the initial mass function'', 2001, MNRAS, 322, 231
\bibitem[\protect\citeauthoryear{Lan et al.}{2024}]{Lan2024} Lan, T.-W., Prochaska J. X., Moustakas J., Siudek, M., Aguilar, J., Ahlen, S., Bianchi, D., Brooks, D., Claybaugh, T., Cole, S., et al., ``DESI Emission-line Galaxies: Unveiling the Diversity of [O II] Profiles and Its Links to Star Formation and Morphology'', 2024, ApJ, 977, 225
\bibitem[\protect\citeauthoryear{Maiolino \& Mannucci}{2019}]{Maiolino2019} Maiolino, R., Mannucci, F., ``De re metallica: the cosmic chemical evolution of galaxies'', 2019, A\&ARv, 27, 3
\bibitem[\protect\citeauthoryear{Mannucci et al.}{2010}]{Mannucci2010} Mannucci, F., Cresci, G., Maiolino, R., Marconi, A., Gnerucci, A., ``A fundamental relation between mass, star formation rate and metallicity in local and high-redshift galaxies'', 2010, MNRAS, 408, 2115
\bibitem[\protect\citeauthoryear{Marino et al.}{2013}]{Marino2013} Marino, R. A., Rosales-Ortega, F. F., S\'anchez, S. F., Gil de Paz, A., V\'ilchez, J., Miralles-Caballero, D., Kehrig, C., P'erez-Montero, E., Stanishev, V., Iglesias-P\'aramo, J., et al., ``The O3N2 and N2 abundance indicators revisited: improved calibrations based on CALIFA and $T_e$-based literature data'', 2013, A\&A, 559, A114
\bibitem[\protect\citeauthoryear{Moustakas \& Kennicutt}{2006}]{Moustakas2006} Moustakas, J., Kennicutt, R. C. Jr., ``An Integrated Spectrophotometric Survey of Nearby Star-forming Galaxies'', 2006, ApJS, 164, 81
\bibitem[\protect\citeauthoryear{O'Donnell}{1994}]{ODonnell1994} O'Donnell, J. E., ``R v-dependent Optical and Near-Ultraviolet Extinction'', 1994, ApJ, 422, 158
\bibitem[\protect\citeauthoryear{Pietrinferni et al.}{2004}]{Pietrinferni2004} Pietrinferni, A., Cassisi, S., Salaris, M., Castelli, F., ``A Large Stellar Evolution Database for Population Synthesis Studies. I. Scaled Solar Models and Isochrones'', 2004, ApJ, 612, 168
\bibitem[\protect\citeauthoryear{Pietrinferni et al.}{2006}]{Pietrinferni2006} Pietrinferni, A., Cassisi, S., Salaris, M., Castelli, F., ``A Large Stellar Evolution Database for Population Synthesis Studies. II. Stellar Models and Isochrones for an $\alpha$-enhanced Metal Distribution'', 2006, ApJ, 642, 797
\bibitem[\protect\citeauthoryear{Pucha et al.}{2025}]{Pucha2025} Pucha, R., Juneau S., Dey A., Siudek, M., Mezcua, M., Moustakas, J., BenZvi, S., Hainline, K., Hviding, R., Mao, Y.-Y., et al., ``Tripling the Census of Dwarf AGN Candidates Using DESI Early Data'', 2025, ApJ, 982, 10
\bibitem[\protect\citeauthoryear{Pettini \& Pagel}{2004}]{Pettini2004} Pettini, M., Pagel, B. E. J., ``[OIII]$/$[NII] as an abundance indicator at high redshift'', 2004, MNRAS, 348, L59
\bibitem[\protect\citeauthoryear{Rong \& Liu}{2026}]{Rong26a} Rong, Y., Liu, S., ``Stellar Surface Density Modulates MgII Cool-gas Outflow Absorption in DESI Star-forming Galaxies'', 2026, eprint~arXiv:2606.28874
\bibitem[\protect\citeauthoryear{Rong et al.}{2026}]{Rong26b} Rong, Y., Liu, S., Zou, H., ``A DESI Calibration of the [O II]--[S II] Electron-density Offset in Integrated Star-forming Galaxies'', 2026, eprint~arXiv:2606.28129
\bibitem[\protect\citeauthoryear{S{\'a}nchez et al.}{2012}]{Sanchez2012} S{\'a}nchez, S. F., Kennicutt, R. C., Gil de Paz, A., van de Ven, G., V\'ilchez, J. M., Wisotzki, L., Walcher, C. J., Mast, D., Aguerri, J. A. L., Albiol-P\'erez, S., et al., ``CALIFA, the Calar Alto Legacy Integral Field Area survey. I. Survey presentation'', 2012, A\&A, 538, A8
\bibitem[\protect\citeauthoryear{S{\'a}nchez et al.}{2014}]{Sanchez2014} S{\'a}nchez, S. F., Rosales-Ortega, F. F., Iglesias-P\'aramo, J., Moll\'a, M., Barrera-Ballesteros, J., Marino, R. A., P\'erez, E., S\'anchez-Blazquez, P., Gonz\'alez Delgado, R., Cid Fernandes, R., et al., ``A characteristic oxygen abundance gradient in galaxy disks unveiled with CALIFA'', 2014, A\&A, 563, A49
\bibitem[\protect\citeauthoryear{Schlegel, Finkbeiner \& Davis}{1998}]{Schlegel1998} Schlegel, D. J., Finkbeiner, D. P., Davis, M., ``Maps of Dust Infrared Emission for Use in Estimation of Reddening and Cosmic Microwave Background Radiation Foregrounds'', 1998, ApJ, 500, 525
\bibitem[\protect\citeauthoryear{Silber et al.}{2023}]{Silber2023} Silber, J. H., Fagrelius, P., Fanning, K., Schubnell, M., Aguilar, J. N., Ahlen, S., Ameel, J., Ballester, O., Baltay, C., Bebek, C., et al., ``The Robotic Multiobject Focal Plane System of the Dark Energy Spectroscopic Instrument (DESI)'', 2023, AJ, 165, 9
\bibitem[\protect\citeauthoryear{Sui et al.}{2026}]{Sui2026} Sui, J., Zou, H., Scholte, D., Saintonge, A., Mezcua, M., Siudek, M., Li, W., Guo, W.-J., Liu, S., Xiao, Y., et al., ``Extremely Metal-poor Galaxies in DESI DR1: Connections to Galaxies in the Early Universe'', 2026, AJ, 171, 351
\bibitem[\protect\citeauthoryear{Tremonti et al.}{2004}]{Tremonti2004} Tremonti, C. A., Heckman, T. M., Kauffmann, G., Brinchmann, J., Charlot, S., White, S. D. M., Seibert, M., Peng, E. W., Schlegel, D. J., Uomoto, A., Fukugita, M., Brinkmann, J., ``The Origin of the Mass-Metallicity Relation: Insights from 53,000 Star-forming Galaxies in the Sloan Digital Sky Survey'', 2004, ApJ, 613, 898
\bibitem[\protect\citeauthoryear{Vazdekis et al.}{2016}]{Vazdekis2016} Vazdekis, A., Koleva, M., Ricciardelli, E., R\"ock, B., Falc\'on-Barroso, J., ``UV-extended E-MILES stellar population models: young components in massive early-type galaxies'', 2016, MNRAS, 463, 3409
\bibitem[\protect\citeauthoryear{York et al.}{2000}]{York2000} York, D. G., Adelman, J., Anderson, J. E., Jr., Anderson, S. F., Annis, J., Bahcall, N. A., Bakken, J. A., Barkhouser, R., Bastian, S., Berman, E., et al., ``The Sloan Digital Sky Survey: Technical Summary'', 2000, AJ, 120, 1579
\bibitem[\protect\citeauthoryear{Zhang et al.}{2017}]{Zhang2017} Zhang, K., Yan, R., Bundy, K., Bershady, M., Haffner, L. M., Walterbos, R., Maiolino, R., Tremonti, C., Thomas, D., Drory, N., et al., ``SDSS-IV MaNGA: the impact of diffuse ionized gas on emission-line ratios, interpretation of diagnostic diagrams and gas metallicity measurements'', 2017, MNRAS, 466, 3217
\bibitem[\protect\citeauthoryear{Zinchenko et al.}{2024}]{Zinchenko2024} Zinchenko, I. A., Sobolenko, M., V{\'i}lchez, J. M., Kehrig, C., ``Characterizing chemical abundance ratios in extremely metal-poor star-forming galaxies in DESI EDR'', 2024, A\&A, 690, A28
\bibitem[\protect\citeauthoryear{Zou et al.}{2024}]{Zou2024} Zou, H., Sui, J., Saintonge, A., Scholte, D., Moustakas, J., Siudek, M., Dey, A., Juneau, S., Guo, W., Canning, R., et al., ``A Large Sample of Extremely Metal-poor Galaxies at z $<$ 1 Identified from the DESI Early Data'', 2024, ApJ, 961, 173
\end{thebibliography}
\end{document}